\documentclass[acmsmall,screen]{acmart}

\AtBeginDocument{%
  }

\setcopyright{acmlicensed}
\copyrightyear{2026}
\acmYear{2026}
\acmDOI{XXXXXXX.XXXXXXX}
\acmConference[Conference acronym 'XX]{Make sure to enter the correct
 conference title from your rights confirmation email}{June 03--05,
  2026}{Woodstock, NY}
\acmISBN{978-1-4503-XXXX-X/2018/06}

\usepackage{enumitem}
\usepackage{lipsum}
\usepackage{array}
\usepackage{wrapfig}
\usepackage{subcaption}
\usepackage{pgfplots}
\usepgfplotslibrary{groupplots}
\pgfplotsset{compat=1.17}
\usepackage{amsmath,amsfonts}
\usepackage[noend]{algpseudocode}
\usepackage{graphicx}
\usepackage{textcomp}
\usepackage{float}
\usepackage{listings}
\usepackage{xspace}
\usepackage{multirow}
\usepackage{amsthm}

\usepackage{balance}
\usepackage{algorithm}
\usepackage{algpseudocode}
\usepackage{colortbl}
\usepackage{subcaption}

\usepackage[skins]{tcolorbox}
\usepackage{xcolor}
\usepackage{multicol}
\usepackage{soul}
\usepackage{framed}
\usepackage{booktabs}

\usepackage{MnSymbol}

\usepackage{booktabs}
\usepackage{multirow}
\usepackage{siunitx}
\newcommand*\colourcheck[1]{%
	\expandafter\newcommand\csname #1check\endcsname{\textcolor{#1}{\ding{52}}}%
}
\colourcheck{blue}
\colourcheck{green}
\colourcheck{red}

\definecolor{custom-blue}{rgb}{0,0,1}

\newtcolorbox{boxB}[2][]{%
  enhanced,colback=white,colframe=black,coltitle=black,
  sharp corners,
  toprule=1.0pt,
  rightrule=0.3pt,
  leftrule=0pt,
  bottomrule=0pt,
  fonttitle=\itshape\scshape\large,
  left=0pt,right=5pt,top=5pt,bottom=3pt,
  attach boxed title to top right={yshift=-0.3\baselineskip-0.4pt,xshift=-5mm},
  boxed title style={tile,size=minimal,left=0.2mm,right=0.5mm,
    colback=white,before upper=\strut},
  title=#2,#1,
}

\newtcolorbox{promptbox}[1][]{ 
  colback=lightgray!10,  
  colframe=gray,      
  fonttitle=\bfseries,    
  fontupper=\tiny,
  title={\tiny #1},
  before=\definecolor{royalblue}{rgb}{0.2549, 0.4118, 0.8824} 
}
\newcommand{\tool}{\textsc{IPA}\xspace}

\newboolean{showcomments}
\setboolean{showcomments}{true}
\ifthenelse{\boolean{showcomments}}
 { \newcommand{\mynote}[2]{
      \fbox{\bfseries\sffamily\scriptsize#1}
        {\small$\blacktriangleright$\textsf{\emph{#2}}$\blacktriangleleft$}}}
        { \newcommand{\mynote}[2]{}}

\usepackage{tikz}

\newtheorem{Key Idea}{Key Idea}

\newcolumntype{L}[1]{>{\raggedright\arraybackslash}p{#1}}

\newcommand{\code}[1]{{\footnotesize\texttt{#1}}}
\usepackage{amsthm}
\definecolor{dkgreen}{rgb}{0,0.6,0}
\definecolor{gray}{rgb}{0.5,0.5,0.5}
\definecolor{lightgray}{rgb}{211, 211, 211}
\definecolor{mauve}{rgb}{0.58,0,0.82}

\definecolor{dkgreen}{rgb}{0,0.6,0}
\definecolor{custom-red}{rgb}{1,0,0}
\definecolor{my-blue}{rgb}{0,0,1}
\definecolor{gray}{rgb}{0.5,0.5,0.5}
\definecolor{mauve}{rgb}{0.58,0,0.82}

\definecolor{c1}{HTML}{f4cccc}
\definecolor{c2}{HTML}{f5cdcd}
\definecolor{c3}{HTML}{fffcfc}
\definecolor{c4}{HTML}{ffffff}
\definecolor{c5}{HTML}{ffffff}
\definecolor{c6}{HTML}{fffdfd}
\definecolor{c7}{HTML}{f5cfcf}
\definecolor{c8}{HTML}{fffbfb}
\definecolor{c9}{HTML}{ffffff}
\definecolor{c10}{HTML}{fffdfd}
\definecolor{c11}{HTML}{fefafa}
\definecolor{c12}{HTML}{fef7f7}
\definecolor{c13}{HTML}{ffffff}
\definecolor{c14}{HTML}{fffefe}
\definecolor{c15}{HTML}{ffffff}
\definecolor{c16}{HTML}{fefafa}
\definecolor{c17}{HTML}{fdf3f3}
\definecolor{c18}{HTML}{fffefe}
\definecolor{c19}{HTML}{fdf5f5}
\definecolor{c20}{HTML}{ffffff}

\begin{document}







\title{Interactive Prompting Assistant: Just-in-Time Requirement Clarification for Agentic Software Development}



\setcopyright{none}

\renewcommand\footnotetextcopyrightpermission[1]{} 

\settopmatter{printacmref=false, printfolios=false}




\author{Cuong Chi Le}
\affiliation{%
 \institution{University of Texas at Dallas}
 \city{Richardson}
 \state{Texas}
 \country{USA}}
\orcid{0009-0006-1322-7396}
\email{cuong.le@utdallas.edu}

\author{Minh Le-Anh}
\affiliation{%
 \institution{FPT Software}
 \city{Hanoi}
 \country{Vietnam}
}
\orcid{0009-0000-2859-7038}
\email{minhla4@fpt.com}

\author{Long Vu}
\affiliation{%
 \institution{FPT Software}
 \city{Hanoi}
 \country{Vietnam}
}
\orcid{0000-0002-5518-4236}
\email{longvdairesearch@gmail.com}

\author{Tien N. Nguyen}
\affiliation{%
 \institution{University of Texas at Dallas}
 \city{Richardson}
 \state{Texas}
 \country{USA}}
\orcid{0009-0006-7962-6090}
\email{tien.n.nguyen@utdallas.edu}




\begin{abstract}
As LLM-based software agents increasingly generate code from
natural-language requirements, developers need assistance not only in
writing code, but also in writing the prompts that specify what the
code should do. We introduce the \emph{Interactive Prompting Assistant
(IPA)}, an IDE-like assistant that provides just-in-time requirement
clarification while a developer is still composing a problem
description. IPA continuously observes the evolving prompt and
selectively surfaces clarification questions when potential omissions
or inconsistencies become sufficiently evident. To provide timely assistance without invoking an expensive reasoning model at every token, IPA separates \emph{when to verify} from
\emph{what to ask}. A lightweight interaction controller processes the
prompt token by token and learns to \textsc{Wait}, \textsc{Track}, or
\textsc{Verify} an emerging concern; a larger reasoning model is
invoked only when verification is warranted to determine whether an
issue is genuine and generate an evidence-grounded question.
We evaluate IPA on 100 HiL-Bench software-engineering tasks containing
353 requirement blockers, using LLMs as simulated prompt authors.
Across three author models, incorporating within-turn IPA before a
matched post-hoc clarification stage improves blocker response by
4.5--13.0 percentage points over post-hoc clarification alone.
Blocker-level transitions further show that IPA acts particularly
strongly as a contextual reminder: recovery of previously omitted
requirements occurs 2.3--5.0$\times$ more often than newly activated
repairs of ambiguities or contradictions. These results demonstrate
the complementary value of progressive, within-turn clarification for
prompt authoring in agentic software development.

\end{abstract}

\maketitle

\section{Introduction}

Integrated development environments (IDEs) have long assisted programmers
\emph{while they write code}. Rather than waiting until an implementation is
complete, modern IDEs provide contextual assistance, including code completion, API
recommendations, diagnostics, and suggested edits~\cite{robillard2010recommendation,svyatkovskiy2020pythia}. 



The emergence of LLM-based software agents is changing the artifact through
which humans increasingly direct software development. Agentic software
development systems can plan changes, navigate repositories, edit multiple
files, execute programs, and test their implementations
\cite{yang2024sweagent,wang2025openhands}. Consequently, for many
agent-driven development tasks, a developer expresses the
desired software behavior through a natural-language \emph{prompt}, often in
the form of a problem description, feature request, or set of requirements,
and delegates substantial implementation work to an agent. This shift raises
a natural question: \emph{if traditional IDEs assist humans while they write
code, what should assist humans while they write the prompts from which
software agents generate code?}

Such assistance is particularly important because natural-language
requirements are rarely perfect when first written. Requirements may omit
behavior that the author implicitly assumes, leave choices underspecified, or
contain statements that conflict with one another. Ambiguity,
incompleteness, and inconsistency have long been fundamental concerns in
requirements engineering
\cite{kamsties2000taming,gervasi2019ambiguity}. Their consequences become
especially important in LLM-based development: when an implementation is
generated directly from a problem description, information left implicit in
the prompt must often be \emph{guessed} by the model. 
Recent work has shown increasing interest in detecting and 
repairing ambiguous natural-language requirements
for code generation \cite{mu2024clarifygpt,jia2025repair}.

We therefore envision an {\bf \em interactive prompting assistant} (IPA): an analogue
of an IDE coding assistant for the prompt-writing stage of agentic software
development. Instead of primarily completing the user's natural-language
text, our current focus is to help the user make the intended software behavior
sufficiently explicit and internally consistent. As a developer writes a
problem description, the assistant observes the evolving prompt and identifies
points at which an important requirement appears to be missing,
underspecified, or inconsistent with information already stated. It can then
pose a targeted clarification question, for example, asking which behavior
is intended in a boundary case or which of two conflicting constraints should
take precedence. The user's answer becomes part of the evolving
prompt, allowing the problem description to be corrected before code
generation is committed to an unintended interpretation.

Existing clarification approaches demonstrate the value of asking users
about incomplete or ambiguous requirements, but generally operate after a
complete requirement or conversational turn has been submitted. For example, a submitted requirement is analyzed and a human is asked clarification questions before generating code, while recent benchmarks evaluate clarification through
multi-turn question-answer interaction~\cite{mu2024clarifygpt,fang2026clarifycodebench}. These approaches address the important
problem of \emph{what} should be clarified, but leave a complementary
interaction point underexplored: assisting the author \emph{while the
requirement itself is still being composed}.

To be complementary to the post-hoc approach, we advocate \emph{progressive requirement clarification}, in which issues are addressed as they become sufficiently evident. We realize this principle through \emph{just-in-time clarification}: continuously observing the evolving prompt while selectively invoking deeper verification at appropriate points during authoring. 
The central research problem is how to interact with a human during his/her prompting for clarification questions.

To address that, we propose a two-model architecture, which separates the
problem of \emph{when to verify} from the more expensive problems of
\emph{whether a genuine requirement issue exists} and \emph{what should be
asked}. A lightweight \emph{interaction model} continuously processes the
evolving prompt and determines whether to \textsc{wait}, \textsc{track}, or
\textsc{verify}. \textsc{Wait} indicates that the current prefix provides
insufficient evidence of an issue; \textsc{track} indicates that a potential
issue has begun to emerge but more context is needed; and \textsc{verify}
indicates that sufficient evidence has accumulated to justify invoking a
larger \emph{verification model}. The interaction model therefore does not
itself determine the missing requirement, resolve an inconsistency, or
generate a clarification question. Instead, it acts as an online controller
that decides \emph{when deeper semantic reasoning becomes worthwhile}.

To keep interaction inexpensive, the controller uses frozen Gemma~4~12B
token embeddings rather than running the full transformer during authoring.
A lightweight GRU with causal attention tracks the evolving prefix, while
separate issue and verification heads estimate whether a concern is emerging
and whether enough evidence has accumulated to invoke the verifier. A small
stateful automaton constrains transitions among \textsc{Wait},
\textsc{Track}, and \textsc{Verify}.
When the interaction model emits \textsc{verify}, the larger verification
model examines the current problem-description prefix to determine whether the
suspected omission, underspecification, or inconsistency is genuine and, when
appropriate, generates a targeted clarification question. After the human
responds or revises the description, the updated prompt re-enters the
incremental interaction process. By learning \emph{where in the token stream
deeper verification should be triggered}, our approach turns requirement
clarification from a post-hoc operation into assistance integrated directly
into prompt authoring. In this sense, the interaction model plays a role
analogous to an IDE mechanism that decides \emph{when} to surface a completion,
warning, or recommendation while a human is writing code.

In summary, this paper makes the following contributions:



\textbf{1. Just-in-time requirement clarification.}
We formulate interactive prompting a new setting for requirements clarification, in which potential
omissions/inconsistencies are identified progressively from an evolving description. 



\textbf{2. Decoupled interaction and verification.}
We introduce a two-layer architecture that separates
low-cost, continuous interaction control from expensive verification and question generation. This enables selective verification without
executing the larger reasoning model at every token.

\textbf{3. Empirical evaluation.} We evaluate {\tool} and compare with
    unassisted, post-hoc, chunked-interaction, and token-level interaction settings. Our evaluation separates downstream authoring outcomes, clarification-question quality, and intervention timing, allowing us to
    study both the benefits and limitations of within-turn clarification.



\section{Motivation}
\label{sec:motiv}

\begin{figure*}[t]
    \centering
    \includegraphics[width=\linewidth]{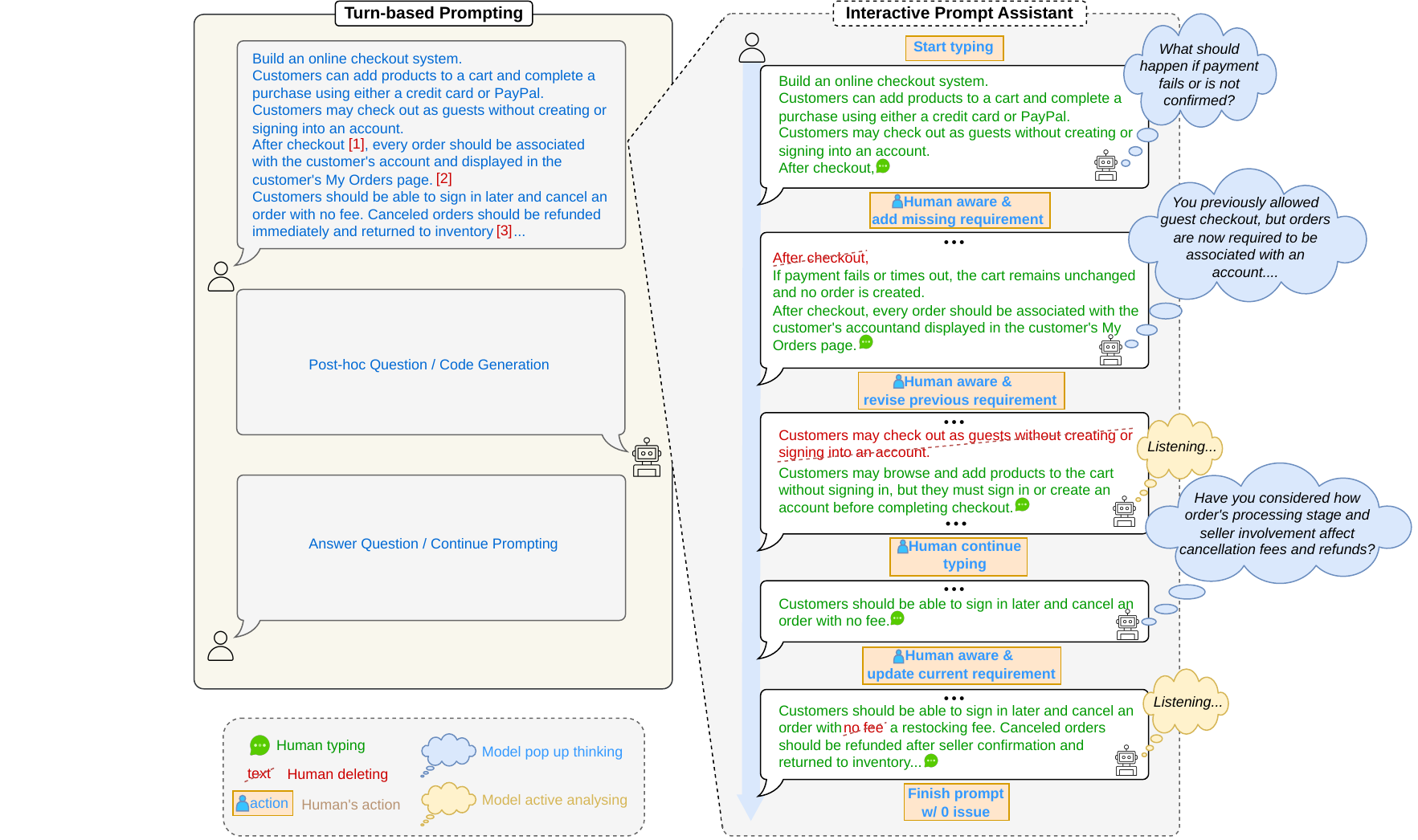}
    \vspace{-18pt}
\caption{Composing a checkout-system prompt under two interaction styles.
\textbf{Left:} turn-based prompting waits until the description is complete.
By then, a missing payment-failure requirement~{\color{red}[1]}, a
guest-checkout versus account-bound-order inconsistency~{\color{red}[2]},
and an implicit no-fee cancellation assumption~{\color{red}[3]} have all
accumulated.
\textbf{Right:} IPA operates within the human turn and surfaces each issue as soon as the evolving prefix
makes it evident. After these in-situ revisions, the prompt is completed
with no remaining issues.}
    \label{fig:motiv}
\end{figure*}

Consider a developer incrementally composing a requirement for an online
checkout system, as shown in Fig.~\ref{fig:motiv}.
Under \emph{interactive prompting} (right), IPA observes the prefix as it is
typed and can surface a lightweight question as soon as an issue becomes
evident. 

{\bf \em \underline{1. IPA as a reminder of missing requirements}.}
At the point {\color{red}[1]} in the prompt in Fig.~\ref{fig:motiv} (Left), nothing already written is incorrect.
The developer has specified payment methods and begun describing checkout, but
has not said what should happen if payment fails.
IPA surfaces a reminder:

\begin{quote}
\small
\emph{What should happen if payment fails or is not confirmed?}
\end{quote}

Still in the context of checkout, the developer adds:

\begin{quote}
\small
{\em If payment fails or times out, the cart remains unchanged and no order is
created.}
\end{quote}

IPA here is a \emph{timely reminder} of a boundary case that is easy to overlook
while the prompt still focuses on the successful usage flow.
Because the question appears while that design context is still salient, the
missing behavior can be added with a local edit.
If the same reminder is deferred until the end, the developer must
return to an earlier topic after attention has already moved on.
Later requirements may then silently assume that payment always succeeds, and
an implementation generated from the incorrect, yet completed prompt must guess the failure
behavior.

{\bf \em \underline{2. IPA as early recognition of a root inconsistency}.}
A different situation arises at {\color{red}[2]}.
The developer has allowed guest checkout, but then requires every order to be
associated with the customer's account and shown in \emph{My Orders}.
These two statements cannot both hold.
When this first clash appears, it is still local, so both the issue and its
root cause---whether a purchaser must have an account---are easy to see.
IPA can intervene as soon as the clash is sufficiently evident: 
\begin{quote}
\small
\emph{You previously allowed guest checkout, but orders are now required to be
associated with an account. Should customers be required to sign in or
create an account before completing checkout?}
\end{quote}
The developer revises the earlier requirement to: {\em "Customers may browse and add products to the cart without signing in, but they must sign in or create an account before completing checkout."}


This early correction does more than repair the visible issue. The remainder of the prompt is written under the corrected assumption that every purchaser has an account, making later account-dependent behaviors naturally consistent rather than further symptoms of the same conflict. Fixing the root early can therefore avoid downstream issues.



{\bf \em \underline{3. IPA as correction of an implicit assumption the
developer still treats as correct}.}
Not every problematic statement is an omission or a contradiction with prior
text.
After the guest-checkout decision has been repaired, a developer continues
and writes that customers may cancel an order with no fee, and that canceled
orders should be refunded immediately ({\color{red}[3]}).
From his view, these statements are still reasonable: they express a
preferred policy, not an error.
If the assumption is left unchallenged, they will keep elaborating the prompt
\emph{as if it were correct}. IPA can intervene while this is still being authored, not by asserting
a contradiction, but by questioning the implicit~assumption:
\begin{quote}
\small
\emph{Have you considered how the order's processing stage and seller
involvement affect cancellation fees and refunds?}
\end{quote}
The developer then updates the \emph{current} prompt: 
{\em "Customers should be able to sign in later and cancel an order with a
restocking fee. Canceled orders should be refunded after seller confirmation
and returned to inventory\ldots"}.
%
%
Subsequent text is then written under a corrected policy rather than under an
unexamined default.
Without such as a reminder, the author will
not self-correct, as (s)he does not recognize any issue.
In the interactive path of Fig.~\ref{fig:motiv}, the correction happens in
place and the developer finishes with no remaining issues.
Without IPA, the biased policy is fully spelled out before any question is asked, and is entangled with {\color{red}[1]} and {\color{red}[2]}.


The three cases are different in kind---{\bf an omission, a root inconsistency, and
a biased assumption}---but each is cheapest to resolve at the moment it first
becomes evident.
A reminder is useful while the relevant context is still being authored.
A root inconsistency is easiest to localize when only the first clash is
visible, and cheapest to repair before later requirements are written on top
of it.
An implicit assumption is most effectively challenged before the developer
has elaborated it into further statements that still look correct from their
point of view.
This demonstrates the benefit of an IPA: observes the prompt \emph{as it is written}
and decides \emph{when} a question is worth asking.
\section{Interactive Prompting Assistant ({\tool})}
\label{sec:approach}

\begin{figure}[t]
\centering
\includegraphics[width=0.7\columnwidth]{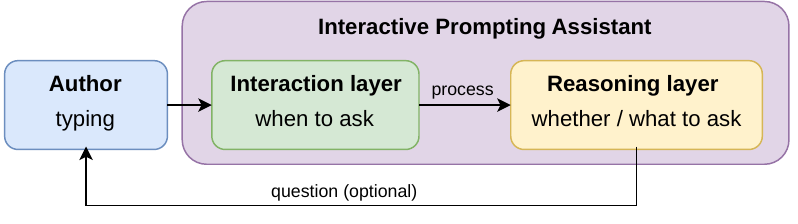}
\vspace{-12pt}
\caption{Two-layer IPA Architecture.
The interaction layer stays with the author as they type and either passes
or calls reasoning on the current prefix.
Reasoning may ask a question; the author may revise; interaction continues
on the updated text.}
\label{fig:arch}
\end{figure}

IPA assists a developer while a prompt is being written. The focused part is
not only what to ask, but when. Asking too late allows issues to accumulate,
whereas invoking semantic reasoning after every token is unnecessarily
expensive and risks premature interaction. We thus separate IPA into
an interaction layer and a reasoning layer (Fig.~\ref{fig:arch}).
The lightweight interaction layer continuously follows the evolving prompt
and decides when deeper analysis is warranted. Only then is the larger
reasoning model invoked to decide whether an issue is genuine
and, if so, what clarification to ask.

\subsection{Two-Layer Architecture}
\label{sec:arch}

Let the prompt be an online stream $x_1,x_2,\ldots$.
At time $t$ only the prefix $p_{1:t}=x_1\cdots x_t$ exists (or $p_t$ for short).
That prefix is incomplete: a gap that looks missing now may be written at
$t{+}k$, and a clash may appear only after the next sentence.
The interaction layer is an online controller:
\begin{equation}
\label{eq:interaction}
I(p_t,s_{t-1})=(d_t,s_t),
\qquad
d_t\in\{\mathrm{hold},\mathrm{process}\},
\end{equation}
where $p_t$ is the current prompt, $s_t$ is the
controller's internal state, and $d_t$ determines
whether deeper verification should be invoked.

Section~\ref{sec:token} refines this binary interface into
\textsc{Wait}, \textsc{Track}, and \textsc{Verify}, where
\textsc{Wait} and \textsc{Track} both correspond to
$d_t=\textit{hold}$ and \textsc{Verify} to
$d_t=\textit{process}$.

A \emph{hold} decision allows authoring to continue
without calling the reasoning layer. A \emph{process}
decision invokes the reasoning layer on the current
prompt:
\begin{equation}
q_t =
\begin{cases}
R(p_t), & d_t=\mathrm{process},\\
\varepsilon, & d_t=\mathrm{hold}.
\end{cases}
\end{equation}
The reasoning layer returns either a clarification
question or $\varepsilon$, indicating that no question
is warranted. A question is displayed only when
$q_t\neq\varepsilon$. This separation distinguishes three events:
recognizing a potential issue, invoking semantic
verification, and displaying a clarification question.
The first two need not interrupt the author; only the
third creates a visible interaction.
When the author revises an earlier portion of the
prompt, the controller state is updated to reflect
the revised text rather than continuing from the
old prefix. Interaction then resumes using the
updated prompt.



\begin{Key Idea}[Interaction and reasoning]
While someone is writing, a good partner does two different things.
They follow the utterance, and they ask only when
a question would help.
IPA splits those stances.
Interaction decides \emph{when} to look; reasoning decides \emph{whether}
to ask and \emph{what}.
\end{Key Idea}

The rest of this section instantiates $I$.
We first describe a simple chunked controller in the spirit of current
turn-like assistance, then show why surface-form chunks do not match how
ideas are produced, and finally propose a token-level controller that
listens as a human listener would.

\subsection{Interaction Layer}
\label{sec:interact}

\begin{figure*}[t]
\centering
\includegraphics[width=0.67\textwidth]{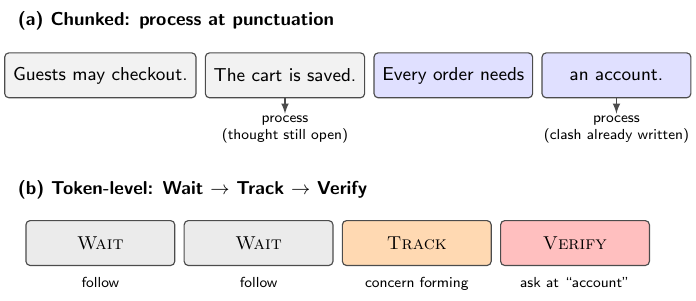}
\vspace{-12pt}
\caption{Two ways to implement the interaction layer on the same prefix.
\underline{(a)}~Chunking waits for a surface-complete span, as current
clarification pipelines wait for a finished requirement or turn, then
calls reasoning---possibly while an idea is unfinished, or after a clash
is already written.
\underline{(b)} Token-level interaction follows the evolving prefix, tracks an emerging
concern, and invokes reasoning once sufficient evidence has accumulated.}
\label{fig:interact}
\end{figure*}

Eq.~\eqref{eq:interaction} only requires a decision $d_t\in\{\mathrm{hold},\mathrm{process}\}$.
How that decision is made is the design of the interaction layer.
We consider two controllers that share the same reasoning layer $\mathrm{R}$.
The first is a simple, training-free schedule that invokes reasoning at
surface-form boundaries. The second is our learned token-level controller,
which continuously observes the evolving prefix and invokes reasoning only
when the available evidence indicates that a potential concern has become
sufficiently mature for verification. Fig.~\ref{fig:interact} contrasts
the two strategies.

\subsubsection{Chunked Interaction}
\label{sec:chunk}

Current clarification methods already decide \emph{when} in a coarse way:
they wait for a finished requirement, then analyse it.
ClarifyGPT takes a complete problem statement, tests whether it is
ambiguous, and only then asks the user~\cite{mu2024clarifygpt}.
Repair of ambiguous descriptions~\cite{jia2025repair} and multi-turn
benchmarks such as ClarifyCodeBench~\cite{fang2026clarifycodebench}
likewise treat a user \emph{turn}---a submitted requirement or
message---as the unit that is ready to be examined.
The shared assumption is: once a surface-complete span has arrived, it is
time to reason.

Chunked interaction is that assumption, moved into the typing loop.
Instead of waiting for the whole prompt, we wait for a small span of
prose---a reproducible proxy for ``a finished thought.''
We segment $p_{1:t}$ into units at sentence-final punctuation or a blank
line (abbreviations and file suffixes are not sentence ends).
A unit longer than $W$ words is split every $W$ words, so a
run-on paragraph cannot defer looking forever.
Let $b_t$ indicate whether a chunk ends at token $t$,
and let $N_t=\sum_{i=1}^{t}b_i$ 
denote the number of completed chunks. Given a
checkpoint stride $K$, the chunked controller is
\begin{equation}
d_t =
\begin{cases}
\mathrm{process},
& b_t=1 \land N_t \bmod K=0,\\
\mathrm{process},
& \text{if prompt submission occurs at }t,\\
\mathrm{hold},
& \text{otherwise}.
\end{cases}
\label{eq:chunk}
\end{equation}
The controller invokes the same reasoning layer as
IPA. Thus, the two approaches differ in their
verification schedules rather than their semantic
verification capabilities.


\paragraph{Why chunking is not enough.}
A period is not the end of an idea.
The author may have closed a sentence while the thought---a condition, an
exception, the other half of a policy---is still being formed.
Chunking does not see that: Eq.~\eqref{eq:chunk} fires because a surface
span ended, so reasoning is asked to analyse a prefix the human does not
yet consider finished (the first $\mathrm{process}$ in
Figure~\ref{fig:interact}a).
The dual failure is just as important.
An omission or contradiction can become locally visible \emph{inside} a
sentence, before the next chunk boundary.
Until that boundary, interaction holds, and the author may keep writing
under the unexamined assumption (the clash at ``account'' in
Fig.~\ref{fig:interact} is visible before the second checkpoint).
In short, chunks follow punctuation, not meaning.
They are simple and reproducible, and they reflect how current systems wait for a completed turn~\cite{mu2024clarifygpt,jia2025repair,fang2026clarifycodebench}, but they do not capture \emph{when the evolving prompt contains sufficient evidence to justify deeper verification}.


\subsubsection{Token-Level Interaction}
\label{sec:token}

\begin{figure*}[t]
\centering
\includegraphics[width=0.85\textwidth]{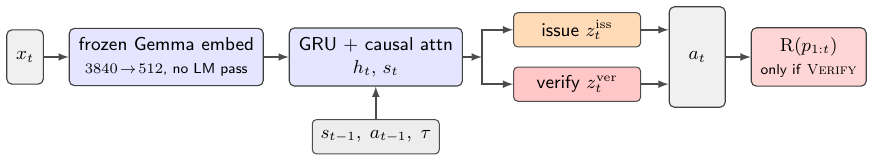}
\vspace{-9pt}
\caption{Token-level controller.
Each token's frozen Gemma embedding is encoded by a GRU with causal
attention. The two heads choose $a_t$, and reasoning runs only on
\textsc{Verify}.}
\label{fig:controller}
\end{figure*}

We therefore let interaction run on every token, in step with typing
(Figure~\ref{fig:interact}b).
The analogue is ordinary conversation: one person speaks, the other
\emph{hears proactively}---they follow each word, notice a concern as it
starts to form, keep listening until the concern is clear enough, and only
then take a turn.
They do not wait for a paragraph break, and they do not interrupt at the
first unclear word.
Figure~\ref{fig:controller} is the network we train for that policy.

The controller chooses among three actions
$a_t \in \{\textsc{Wait}, \textsc{Track},
\textsc{Verify}\}$:

\begin{itemize}
    \item \textsc{Wait}: The available prefix does not
    currently provide sufficient evidence of a concern.
    No reasoning call is made ($d_t=\mathrm{hold}$).

    \item \textsc{Track}: A potential concern is emerging,
    but the available evidence is not sufficient
    to justify deeper verification. The controller
    continues observing the evolving prompt
    ($d_t=\mathrm{hold}$).

    \item \textsc{Verify}: The accumulated evidence
    warrants invoking the reasoning model on the
    current prompt prefix ($d_t=\mathrm{process}$).
\end{itemize}

Neither \textsc{Wait} nor \textsc{Track} produces a
user-visible interaction. A \textsc{Verify} action
invokes the reasoning model, which may either return
a clarification question or remain silent
($q_t=\varepsilon$). Consequently, verification
and interruption are separate events. Multiple concerns may therefore
contribute to the controller's decision to invoke
verification. When verification is triggered, the reasoning layer
examines the available prefix and selects at most one
sufficiently grounded clarification question.

\begin{figure}[t]
\centering
\vspace{-9pt}
\includegraphics[width=0.53\columnwidth]{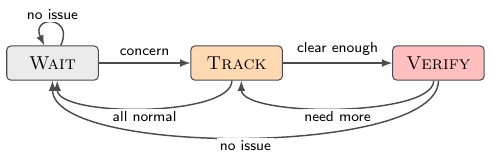}
\vspace{-15pt}
\caption{Allowed transitions of the token-level controller.}
\label{fig:automaton}
\end{figure}

\paragraph{State-transition constraints.}
Figure~\ref{fig:automaton} constrains the action sequence so the
controller does not jump straight to a question or ask twice in a row.
From \textsc{Wait}, it may stay when there is no issue, or enter
\textsc{Track} on a concern. From \textsc{Track}, it returns to
\textsc{Wait} when the prefix is normal, or enters \textsc{Verify} when
the evidence is clear enough. After \textsc{Verify}, it returns to
\textsc{Wait} if no issue, or to \textsc{Track} when more
detail is needed. Illegal transitions are masked during
training and inference:
\begin{equation}
  a_t \;=\; \arg\max_{a \in \mathcal{A}(a_{t-1})}
  \ell_t(a),
\label{eq:action}
\end{equation}
where $\mathcal{A}(\cdot)$ is the allowed-next set in
Figure~\ref{fig:automaton} and $\ell_t$ are the action logits below.

\paragraph{Architecture.}
Calling a 12B transformer on every keystroke would defeat the layer split.
Each token $x_t$ is represented by a frozen Gemma~4~12B embedding
$e_t\in\mathbb{R}^{3840}$, looked up from the embedding table (no LM
forward pass).
A projection $e_t\mapsto \tilde{e}_t\in\mathbb{R}^{512}$ is consumed by a
two-layer GRU with causal attention, yielding a state $h_t$ that may use
$p_{1:t}$ but not the future:
\begin{equation}
  h_t,\,s_t \;=\; \mathrm{Enc}(\tilde{e}_t,\,s_{t-1}).
\label{eq:enc}
\end{equation}
Two heads share $h_t$.
The issue head $z^{\mathrm{iss}}_t$ asks whether a concern is becoming
visible (leave \textsc{Wait} for \textsc{Track}).
The verify head $z^{\mathrm{ver}}_t$ asks whether that concern is mature
enough to process.
We assemble
\begin{equation}
\begin{aligned}
  \ell_t(\textsc{Wait})   &= -z^{\mathrm{iss}}_t,\\
  \ell_t(\textsc{Track})  &=  z^{\mathrm{iss}}_t,\\
  \ell_t(\textsc{Verify}) &=  z^{\mathrm{iss}}_t + z^{\mathrm{ver}}_t,
\end{aligned}
\label{eq:logits}
\end{equation}
We then apply the transition mask in Eq.~(4). The issue head controls the
relative evidence for remaining in \textsc{Wait} versus tracking a concern,
while the verification head increases the score of \textsc{Verify} relative
to \textsc{Track}. Because \textsc{Verify} is reachable only from
\textsc{Track}, verification requires the controller to have entered the
tracking state on a preceding token. The previous action $a_{t-1}$ and track
age $\tau_t$ additionally condition the action prediction, allowing the
controller to distinguish a newly emerging signal from a concern that has persisted as the prefix grows.


\subsubsection{Interaction Policy}
\label{sec:learning-policy}

The interaction controller must distinguish between the point at which a
potential requirement concern first emerges and the later point at which
the available evidence is sufficient to justify semantic verification.
We supervise the controller using two token positions for each
annotated issue $j$: the \emph{concern onset}
$t_j^{\mathrm{onset}}$ and the \emph{verification point}
$t_j^{\mathrm{verify}}$, where
$t_j^{\mathrm{onset}} \leq t_j^{\mathrm{verify}}$.
The concern onset marks the first token at which the evolving prompt
provides evidence of a potential issue. The verification point marks the
token at which that concern has become sufficiently clear to warrant
invoking the reasoning layer. These positions induce the token-level
actions used for training:
tokens before $t_j^{\mathrm{onset}}$ are labeled \textsc{Wait};
tokens from $t_j^{\mathrm{onset}}$ up to, but not including,
$t_j^{\mathrm{verify}}$ are labeled \textsc{Track}; and
$t_j^{\mathrm{verify}}$ receives a one-token \textsc{Verify} label.
When multiple concerns are active at the same token, we assign the
highest-priority action, with
$\textsc{Verify} > \textsc{Track} > \textsc{Wait}$.

The two positions differ by issue type. For potentially missing information, onset occurs once the relevant functional context is available, while verification waits until the prefix provides sufficient evidence that the behavior or constraint remains unspecified. For an inconsistency, onset marks the emergence of potentially conflicting requirements, while verification occurs once the prefix provides sufficient evidence of the conflict.


We train the controller on the resulting \textsc{Wait}--\textsc{Track}--
\textsc{Verify} action sequences while applying the transition constraints
of Eq.~(4). Although the training labels are constructed from annotated
authoring traces, the controller at token $t$ receives only the prefix
available through $t$ and cannot access future text. It therefore learns
from the evolving prefix both when a concern begins to emerge and when
sufficient evidence has accumulated to justify deeper semantic reasoning.
Section~\ref{sec:rq2-controller} describes the annotation procedure and evaluates the learned verification timing.

\subsection{Reasoning Layer}
\label{sec:verify}


The interaction layer determines when deeper analysis
is warranted, but it does not determine whether a
requirement issue is genuine or what clarification
should be presented. These responsibilities belong
to the reasoning layer.
When invoked, the reasoning model receives the current
prompt prefix and applicable clarification history.
It does not have access to future prompt content.
Its task is to determine whether the available
evidence supports a useful clarification question.

We distinguish two principal issue categories.

\textbf{\em Potentially missing information.}
The prompt leaves an important behavior, constraint,
or design choice unspecified. Because the prompt is
still being written, absence alone does not establish
an error. The reasoner must determine whether the
available context supports a timely reminder.

\textbf{\em Inconsistent information.}
Two or more requirements express conflicting
constraints or assumptions. The reasoner identifies
the conflicting statements and formulates a question
that allows the developer to clarify the intended
behavior.


\textbf{\em Evidence-grounded clarification.}
For each invocation, the reasoning model returns either
$\varepsilon$, indicating that no sufficiently grounded
clarification is warranted, or a structured clarification
$
q_t = (\kappa_t,E_t,Q_t),
$
where $\kappa_t$ denotes the issue category,
$E_t$ identifies the supporting evidence in the
current prompt, and $Q_t$ is the user-visible
clarification question.
For an inconsistency, the evidence identifies the
conflicting requirements. For potentially missing
information, it identifies the existing functional
context that makes the unspecified behavior relevant.
The reasoner must ground its question in the available
prefix rather than infer that an unstated requirement
is necessarily erroneous. When the evidence is
insufficient, it returns $\varepsilon$.

{\bf \em Revision and continuation.}
A displayed question is nonmodal: the author may answer, dismiss, or defer
it. If the author answers or edits the requirement, the resulting prompt
becomes the new input to the interaction layer, whose state is recomputed
from the revised text as needed. IPA surfaces clarification
opportunities without automatically modifying the author's requirements.

\section{Empirical Evaluation}
\label{sec:exp}
\label{sec:eval}


\textbf{RQ1. [Simulated Authoring]}
How does within-turn clarification affect simulated authors' responses to
requirement blockers, particularly beyond matched post-hoc clarification?

\noindent\textbf{RQ2. [Question Quality in Issue Detection]}
How relevant are the clarification questions generated by IPA
and post-hoc clarification, and how completely do they cover the registered
blockers?


\noindent\textbf{RQ3. [Latency]}
How soon after an issue appears in the prompt can IPA return a
question?


{\bf \em Models.}
We use DeepSeek-V4-Flash, Kimi-K2.5 and Grok 4.1-Fast as
candidate models, GPT-5.6~Luna as the judgement model,
Gemma-4-12B embeddings for the interaction layer
(Section~\ref{sec:token}).

\subsection{LLM-as-Human Simulation (RQ1)}
\label{sec:rq1}

\subsubsection{Dataset}
\label{sec:dataset}

We evaluate on the software-engineering portion of
HiL-Bench~\cite{trinh2026hilbenchhumaninloopbenchmarkagents},
which is constructed from SWE-Bench Pro tasks and introduces
human-validated information gaps that require clarification.
We use all of the eligible 100 publicly available SWE tasks
(\texttt{swe\_0}--\texttt{swe\_99}).
These tasks contain 353 registered blockers, corresponding to
an average of 3.53 blockers per task.

A \emph{blocker} represents information that cannot be safely
resolved by guessing. HiL-Bench distinguishes three types:
\emph{missing information}, where information needed to
determine the intended behavior is absent;
\emph{ambiguous requests}, where multiple plausible
interpretations remain possible; and
\emph{contradictory information}, where two stated requirements
cannot simultaneously be satisfied.

Each blocker $b$ is associated with a structured registry entry
containing its type, a description of the information gap, an
exact intended resolution $r_b$, and example trigger questions.
For example, if HiL-Bench says that the required behavior for a failed payment is genuinely missing, and IPA asks: {\em “What should happen if the payment fails?”}
then the question matches a registered blocker.
The resolution is hidden from the models during normal
authoring and is used only by the experimental environment to
simulate information that would normally reside with the human.
This distinction is important for our study. Asking the simulated author
to invent details that are intentionally absent would not faithfully
model clarification, since the answer should come from the requirement
author rather than from the assistant. We thus treat the registered
HiL-Bench resolution $r_b$ as a controlled proxy for the author's latent
intent. It remains hidden during normal authoring and is revealed only
after a clarification question has successfully targeted the corresponding
blocker.

\subsubsection{LLM-Based Author Simulation}
\label{sec:simulation}

We separate four {\bf model roles} in the experimental pipeline.

The {\bf \em author model} $M_A$ plays the role of a developer writing
and revising the problem description. We instantiate $M_A$ using multiple
model families to reduce dependence on any single simulated author.
The {\bf \em verification model} $M_V$ performs IPA's semantic reasoning.
When invoked, it examines the available problem-description text and
determines whether a clarification question is warranted. The same $M_V$
is used in both the IPA and post-hoc conditions. The {\bf \em blocker oracle} $\mathcal{O}$ is an evaluation-only component backed by the HiL-Bench blocker registry. It determines whether a question
targets a registered blocker and, when appropriate, provides that blocker's
gold resolution. Neither the interaction controller nor $M_V$ has access to
$\mathcal{O}$ when deciding whether or what to ask.
Finally, the {\bf \em judgment model} $M_J$ evaluates the completed problem
descriptions against the gold blockers. It is used only after authoring and
does not participate in the interaction.
Separating these roles prevents direct exposure of the blocker registry to
IPA while allowing the experimental environment to provide the simulated
author with the latent information represented by a gold resolution.



\vspace{1pt}
{\em Why Simulate Authoring with an LLM?} 
RQ1 requires controlled, paired authoring trajectories in which the same
initial prompt is exposed to different clarification conditions and,
after an intervention, regenerated from the same earlier prefix. Such
counterfactual control is difficult with human participants because once
a person has seen or written a continuation, that knowledge cannot be
removed before replaying another condition. LLM-based authors allow us to
hold the initial text, model, decoding configuration, and latent resolution
fixed while varying the interaction condition. This simulation is intended
as a controlled evaluation of the interaction mechanism, not as evidence
about human usability or cognitive behavior.

A conventional autonomous coding agent is also unsuitable because its
repository navigation, tool calls, code modification, and execution cannot
be controlled at the granularity needed to study interventions at particular
locations in a textual requirement.

LLM-based simulation provides this control but introduces another limitation:
once a continuation has been generated, current interfaces do not allow us
to erase that suffix, revise an earlier prefix, and resume as if the discarded
continuation had never been produced. We therefore design a
\emph{rewind-and-resume} simulation, which approximates the authoring process
in which a developer receives feedback, revises earlier requirements, and
continues under the updated context.

\subsubsection{Evaluation Procedure}
\label{sec:procedure}

{\em Initial Authoring}. For each defective HiL-Bench problem statement $S$, the author
model $M_A$ is instructed to rewrite the statement as a
developer preparing a problem description for a coding agent.
The author may clarify wording, repair issues it notices, and
add information it believes is missing, but it is not shown the
gold blockers or their resolutions.
The initial rewrite is as follows: $P^{(0)}=(x_1,x_2,\ldots,x_n)$.
For each task and author model, we cache this initial rewrite
and use exactly the same $P^{(0)}$ to initialize all experimental
conditions. Thus, differences between conditions cannot be
attributed to independently sampled initial prompts.
We generate one $P^{(0,r)}$ for each run $r$ and fork that same initial prompt
across all conditions. The resulting unit of paired comparison
is therefore
\emph{author model $\times$ task $\times$ run $\times$ blocker}.

{\bf \em NoAssist}. In \textsc{NoAssist}, the candidate receives no clarification
assistance. The initial rewrite $P^{(0)}$ is treated as the completed problem description.



{\bf \em PostHoc}. In \textsc{PostHoc}, clarification occurs only after the
initial rewrite has been completed. The verification model
$M_V$ receives the entire prompt $P^{(0)}$ and may generate clarification questions.

For an accepted question that targets a registered blocker
$b$, the oracle $\mathcal{O}$ returns the corresponding
resolution $r_b$. This response represents information that
would normally be supplied by the human requirement author.
The author can then use the clarification to revise the prompt.
Questions that do not correspond to a registered blocker
receive no privileged gold answer; if accepted, the author responds
using only the information available in its current context.

The author model may then revise any part of the completed prompt based on
the accepted clarifications. The resulting text is the final PostHoc prompt.


{\bf \em IPA: Rewind-and-Resume Authoring}.
IPA begins from the same initial rewrite $P^{(0)}$, but exposes
it incrementally to the interaction mechanism. The interaction
controller, simulating our interaction model, processes the prompt token by token and decides when
there is sufficient evidence to invoke $M_V$.
Suppose verification is invoked after token $x_t$. The
verification model receives only the currently available prefix
$
P_{\leq t}=(x_1,\ldots,x_t)
$
together with the applicable clarification history. It has no
access to the unobserved suffix
$(x_{t+1},\ldots,x_n)$.

If the verifier returns a question $q_t$, the
author decides whether it is useful based only on the currently
available authoring context. This decision is made before any
gold resolution is revealed.

If the author rejects $q_t$, no revision is forced. The
interaction is recorded as rejected and processing continues
along the current authoring trajectory.

If the author accepts $q_t$, and the question targets a
registered blocker $b$, the experimental oracle provides the
blocker's hidden intended resolution $r_b$. We then
\emph{rewind} the simulated authoring process to the
intervention point $x_t$. All text following $t$ is discarded:
$
(x_{t+1},\ldots,x_n)\leftarrow\varnothing.
$
The author may revise any portion of the existing prefix using
the newly clarified information, producing
$\widetilde{P}_{\leq t}$. It then generates a new continuation
conditioned on that revised prompt:
$
P' =
\widetilde{P}_{\leq t}
\circ
M_A(\widetilde{P}_{\leq t}).
$
Because resumed generation uses a fresh LLM API call, the discarded suffix is omitted from the model context and cannot influence the continuation through conversational memory. This allows the simulation to approximate a clean rewind that would be impossible to guarantee with the same human author, who may still remember the discarded text. Subsequent requirements are therefore generated from the revised context rather than from the obsolete continuation.
After a revision, the interaction-controller states derived from
obsolete text are invalidated and reconstructed from the
updated prompt. Clarification history is retained so that
already resolved issues can be recognized as such during
subsequent verification.
The procedure repeats until the author completes the prompt,
no further clarification is triggered, or the common
interaction budget is exhausted.
Thus, rewind-and-resume approximates the revise-and-continue dynamics of
within-turn clarification while preserving controlled counterfactual
comparison across conditions.


\subsubsection{Controlled Comparison and Metrics}

We aim to compare two authoring
pipelines, \textsc{NoAssist+PostHoc} (or PostHoc for short) and \textsc{IPA+PostHoc},
so that the principal experimental difference is whether
clarification can occur \emph{during} authoring.
PostHoc clarification is a common final stage rather
than a competing approach: both pipelines are allowed to refine
the completed description after authoring, while only
\textsc{IPA+PostHoc} can intervene before completing description.

For each task and author model, the two pipelines share the same initial
authoring process, $M_A$, $M_V$, clarification-response mechanism, revision
instructions, and decoding configuration. Neither model sees the gold blocker
registry before a question is accepted, and in both pipelines a
gold resolution is revealed only after the accepted question is matched to the
corresponding blocker.



The two pipelines therefore differ primarily in the \emph{timing and context}
of clarification. In \textsc{NoAssist+PostHoc}, clarification occurs only after
the problem description is complete. In \textsc{IPA+PostHoc}, $M_V$ can reason
over intermediate prefixes, allowing clarified/revised context to influence subsequent
requirements. Both pipelines then undergo the same PostHoc refinement after
authoring. Thus, improvements on description quality by \textsc{IPA+PostHoc} over
\textsc{NoAssist+PostHoc} reflect the benefit of progressive clarification
during authoring rather than simply receiving more clarification.

We report three complementary metrics.

{\bf \em 1. Blocker Response Rate (BRR).}
Let $B$ be the registered blockers for a task. After authoring, $M_J$ independently assigns each blocker one of three labels:
\code{keep\_same} if the blocker remains unaddressed,
\code{try\_to\_fix} if the author makes an observable attempt to resolve an
issue, and
\code{add\_needed\_info} if previously missing requirement information is
introduced. We define
$
\mathrm{BRR}
=
100\cdot
\frac{
n_{\mathrm{try\_to\_fix}}+
n_{\mathrm{add\_needed\_info}}
}{
n_{\mathrm{blockers}}
}.
$
BRR measures whether the author responds to a blocker, not whether the repair
is correct.

{\bf \em 2. Candidate Acceptance.}
Let $Q$ be the questions posed by a condition and $Q_{\mathrm{acc}}$ those the
simulated author chooses to act upon. We report
$
\mathrm{AcceptRate}
=
\frac{|Q_{\mathrm{acc}}|}{|Q|}.
$
Acceptance reflects the author's perceived usefulness of the question rather
than its correctness against a gold resolution.

{\bf \em 3. Blocker-Level Transitions.}
We pair judgments for the same gold blocker under \textsc{NoAssist} and IPA:
$L_b^{N}\rightarrow L_b^{I}$.
For missing-information blockers,
\code{keep\_same}$\rightarrow$\code{add\_needed\_info} is a
\emph{reminder-driven recovery}, measured as
$\mathrm{\code{ReminderRate}}
= n_{\texttt{keep}\rightarrow\texttt{add}}/n_{\texttt{keep}}^{N}$.
For ambiguous and contradictory blockers,
\code{keep\_same}$\rightarrow$\code{try\_to\_fix} is a {\em repair activation},
measured as
$\mathrm{\code{RepairRate}}
= n_{\texttt{keep}\rightarrow\texttt{fix}}/n_{\texttt{keep}}^{N}$.
These metrics distinguish whether IPA improves by recovering missing requirements, triggering repairs, or both.
\subsubsection{Judgment Protocol}
\label{sec:judgment}

All final prompts from the four authoring conditions are evaluated using the
same judgment model $M_J$ and structured rubric. For each blocker, the judge
receives the original defective statement, the final rewritten statement,
the blocker description, and its gold resolution. Condition identity is
withheld from the judge.

The judge assigns a response label
$L_b\in\{\code{keep\_same},\code{try\_to\_fix},
\code{add\_needed\_info}\}$. For blockers that are addressed, it also
assigns
$C_b\in\{\code{correct},\code{incorrect}\}$,
indicating whether the resulting requirement is consistent with the gold
resolution. Separating $L_b$ from $C_b$ prevents an attempted repair from
being automatically counted as a correct one. 
To assess judgment reliability, we manually audited a stratified sample of
340 blocker-level judgments, with 10 judgments sampled from each combination
of blocker type, author model, and condition. This sample size
provides a 95\% confidence level with a $\pm5$ percentage-point
margin of error for estimating overall agreement. The manual annotations agreed with $M_J$ on
91.7\% of response labels ($\kappa=0.86$) and 94.4\% of correctness labels
($\kappa=0.89$), indicating substantial agreement between the automated and
manual judgments.

\subsubsection{Empirical Results}
\label{sec:rq1-results}

\begin{table*}[t]
  \centering
  \footnotesize
\caption{Blocker response of three simulated authors under four
authoring pipelines on HiL-Bench tasks containing 353 gold
blockers. BRR measures the fraction of gold blockers addressed in
the final description. AcceptRate is the fraction of IPA questions
accepted by the author. ReminderRate measures
\code{keep\_same}$\rightarrow$\code{add\_needed\_info}
transitions among initially unaddressed missing-info blockers;
RepairRate measures
\code{keep\_same}$\rightarrow$\code{try\_to\_fix}
transitions among initially unaddressed ambiguity and contradiction~blockers.}
  \label{tab:rq1}
  \tabcolsep 5pt
  \vspace{-6pt}
  \begin{tabular}{l cccc ccc}
  \toprule
  & \multicolumn{4}{c}{BRR (\%)}
  & \multicolumn{3}{c}{IPA interaction (\%)} \\
  \cmidrule(lr){2-5} \cmidrule(lr){6-8}
  Candidate model
    & \textsc{NoAssist} & \textsc{PostHoc} & IPA & IPA+\textsc{PostHoc}
    & AcceptRate & ReminderRate & RepairRate \\
  \midrule
  DeepSeek-V4-Flash & 15.6 & 62.6 & 37.1 & 75.6 & 93.6 & 29.0 & 10.2 \\
  Grok-4-1-fast     & 23.5 & 75.1 & 29.2 & 79.6 & 98.5 & 19.5 &  8.5 \\
  Kimi-K2.5         & 31.2 & 71.6 & 52.4 & 78.2 & 95.0 & 43.1 &  8.7 \\
  \bottomrule
  \end{tabular}
  \end{table*}



Table~\ref{tab:rq1} reports the results.

\vspace{1pt}
{\bf \em 1. Early clarification provides complementary gains beyond
post-hoc clarification.} IPA and \textsc{PostHoc} operate at different stages of prompting. IPA interacts \emph{within a turn}, while the problem description is still being written, whereas \textsc{PostHoc} operates after the turn is complete. They are therefore complementary: IPA can clarify issues early enough to influence subsequent authoring, while \textsc{PostHoc} addresses issues that remain in the completed description.
This complementarity is reflected in the results.
The consistently higher BRR of IPA+\textsc{PostHoc} over \textsc{PostHoc} alone shows that incorporating within-turn IPA adds value beyond post-hoc clarification by itself.
Relative to \textsc{PostHoc} alone, BRR increases from 62.6\% to 75.6\% for DeepSeek
($+13.0$\,pp), from 75.1\% to 79.6\% for Grok ($+4.5$\,pp), and from 71.6\% to 78.2\% for Kimi
($+6.6$\,pp). In fact, IPA+\textsc{PostHoc} achieves the highest BRR for all three
simulated authors.
Because the two conditions use the same clarification budget, author and
verification models, and post-hoc procedure, the comparison isolates the effect
of using clarification during authoring rather than only after completion.



\vspace{1pt}
{\bf \em 2. Clarification substantially increases blocker response.}
Without clarification, the three simulated authors address only
15.6--31.2\% of the gold blockers: 15.6\% for DeepSeek-V4-Flash,
23.5\% for Grok-4-1-fast, and 31.2\% for Kimi-K2.5.
Thus, many requirement issues remain unaddressed when the author
rewrites the defective statement independently.

Adding clarification improves BRR for every author model.
IPA raises BRR to 37.1\% for DeepSeek ($+21.5$\,pp),
29.2\% for Grok ($+5.7$\,pp), and 52.4\% for Kimi
($+21.2$\,pp).
Post-hoc clarification achieves higher BRR
(62.6--75.1\%) than IPA alone. 
This is consistent with the harder online setting faced by IPA:
IPA observes an evolving prefix and must decide both whether and when
to invoke clarification, whereas \textsc{PostHoc} reasons over the
completed problem description. This result indicates the complementary nature between IPA and PostHoc, consistent with the above results.



\begin{boxB}{Finding 1}
Clarification substantially increases blocker response, and
within-turn IPA complements post-hoc clarification. Under the same
total clarification budget, IPA+\textsc{PostHoc} improves BRR by an
additional $+4.5$ to $+13.0$ percentage points over
\textsc{PostHoc} alone across the three simulated authors. BRR also rises from 15.6--31.2\% under unassisted rewriting to 75.6--79.6\% under IPA+\textsc{PostHoc}. 
\end{boxB}

{\bf \em 3. IPA more frequently recovers missing information than
activates repairs.}
Blocker-level transitions reveal where IPA's gains originate.
Among missing-information blockers left unaddressed under
\textsc{NoAssist}, IPA causes 43.1\% for Kimi, 29.0\% for
DeepSeek, and 19.5\% for Grok to transition from
\code{keep\_same} to \code{add\_needed\_info}.
In contrast, among initially unaddressed ambiguity and
contradiction blockers, only 8.7\%, 10.2\%, and 8.5\%,
respectively, transition to \code{try\_to\_fix}.

Thus, across the three author models, reminder-driven recovery is
consistently higher than newly activated repair, by approximately
$2.3\times$--$5.0\times$. This result supports IPA's role as a
contextual reminder: its strongest effect is prompting simulated
authors to supply detailed requirements that they did not
include independently, rather than inducing additional repairs of
already expressed~conflicts.

{\bf \em 4. IPA questions are frequently accepted by simulated authors.}
The simulated authors accept most IPA clarification questions before
seeing any gold resolution: 93.6\% for DeepSeek, 98.5\% for Grok,
and 95.0\% for Kimi. For example, DeepSeek accepts 249 of 266
questions, while Grok accepts 334 of 339. Rejected questions are
skipped and do not force a revision. Thus, most IPA questions are
judged actionable by the simulated authors before the gold resolution
is revealed.


AcceptRate measures the author's perceived usefulness of a question, not its
objective correctness or selectivity. We therefore interpret it with question
precision, blocker recall, AskF1, and the number of questions posed. A high
AcceptRate indicates that most questions are considered worth acting on, but
does not by itself show that IPA avoids excessive questioning.


AcceptRate and BRR capture complementary outcomes. An author may accept
a question without ultimately changing a registered blocker, or may make
only a partial or ineffective revision. This distinction is visible for
Grok: it has the highest AcceptRate (98.5\%) but the smallest IPA BRR
gain ($+5.7$\,pp). Thus, AcceptRate reflects whether the simulated author
chooses to act on a clarification, whereas BRR measures whether the final
description responds to benchmarked requirement issues.

\vspace{-9pt}

\begin{boxB}{Finding 2}
Among initially unaddressed missing-info blockers, IPA produces
\code{keep\_same} $\rightarrow$ \code{add\_needed\_info} transitions
$2.3\times$--$5.0\times$ more frequently than
\code{keep\_same} $\rightarrow$ \code{try\_to\_fix} transitions among
initially unaddressed ambiguity and contradiction blockers. This suggests
a particularly strong role for IPA as a contextual reminder, while the
two rates concern different blocker populations.
\end{boxB}

\vspace{-12pt}

\begin{boxB}{Finding 3}
Simulated authors accept 93.6--98.5\% of IPA's clarification questions
before receiving any gold resolution. This indicates that most questions
are judged actionable by the simulated authors; objective question quality
and interaction burden are evaluated separately through question precision,
blocker recall, AskF1, and the number of questions posed.
\end{boxB}


\subsection{Question Quality in Issue Detection (RQ2)}
\label{sec:rq2}

RQ1 evaluates IPA from the simulated author's perspective: whether
clarification questions are perceived as useful and whether acting on
them improves the final problem description. Both signals are indirect:
acceptance is author-dependent, while BRR measures the resulting text
rather than the intrinsic quality of the clarification.
RQ2 therefore evaluates the quality of IPA's interactive issue-detection
process, examining how effectively its clarification questions surface
registered blockers and contextually valid concerns during authoring.
%
A question can be useful in two ways. It may target a registered blocker, or it may be a valid \emph{reminder} about {\em the prefix available when it was asked}. We apply this contextual-validity judgment to all
methods: interactive questions are judged against the prefix $P_{\leq t}$ available at the time of interaction, whereas \textsc{PostHoc} questions are judged against the completed problem
description. For interactive methods, a valid non-blocker question is
designated a \emph{prefix-grounded reminder}. Otherwise, a question valid at an intermediate prefix could be incorrectly counted as a false positive once the missing detail is supplied later.

\subsubsection{Compared Methods} As with RQ1, we study the following methods:

\emph{PostHoc} applies the same reasoning layer
after the complete problem description is available and corresponds to the
final clarification stage used in RQ1.

\emph{IPA (Chunked)} follows the surface-form schedule of
Eq.~\eqref{eq:chunk} with stride $K=2$. The learned interaction controller is
disabled, and verification is triggered only at the sentence
checkpoints.

\emph{IPA (Token+Fallback)} uses the learned Wait--Track--Verify controller
(Section~\ref{sec:token}) to determine, token by token, when the current
prefix warrants verification. As a safety fallback, if the controller does
not verify for two consecutive sentences, the chunked schedule triggers
verification at that checkpoint, preventing extended passages from delaying
semantic inspection.


\emph{IPA (Token+Fallback)+PostHoc} applies the token-level interactive
assistant in IPA and then performs the same PostHoc clarification
after the description is complete. This mirrors the paired comparison in RQ1
and measures whether within-turn issue detection provides complementary
coverage beyond a final post-hoc stage.

Within each model setting, all methods share $M_V$, $M_J$, the oracle $\mathcal{O}$,
the question budget, the revision procedure, and the repeat-suppression rule.
They differ only in the verification schedule and, consequently, in how much
of the evolving description is available as clarification is requested.



\subsubsection{Evaluation Procedure}
\label{sec:rq2-protocol}
{\bf \em Streaming and when to verify}. 
Interactive methods parse the description token by token and observe only the current prefix $P_{\leq t}$, while \textsc{PostHoc} receives~the~completed description. At each verification, the reasoning model receives $P_{\leq t}$ and returns one~question. 


{\bf \em What happens when a question is right.} Each generated question goes to the judgment model, which sees the blocker descriptions and decides whether the question asks about any of the blockers (Section~\ref{sec:rq2-judgment}). If it does, the oracle $\mathcal{O}$ returns that blocker's gold resolution $r_b$, and a separate model revises the text written so far to incorporate it, exactly as the accepted questions were applied in RQ1. Streaming then continues over the remaining description from the revised text. Revising immediately prevents an issue from cascading: once a blocker's resolution replaces the prefix, later verification calls analyze text in which that issue no longer exists, avoiding repeated credit or penalty. This also reflects the intended interaction, where a developer answers a question, revises the text, and continues writing. Resolved blockers and prior irrelevant questions are included in later verification prompts to discourage repetition. Each call may return at most one question.



\subsubsection{Question Judgment}
\label{sec:rq2-judgment}

$M_J$ first receives the registered blockers with their gold resolutions and
decides which of them, if any, a question $q$ requests. Semantically equivalent
phrasings count as matches, and one question may cover several blockers.
Following HiL-Bench, a blocker can be credited to at most two questions, so
restating a successful question earns nothing. If $q$ matches no blocker and was
asked during streaming, a second judgment sees only the prefix $P_{\leq t}$ that
was available at the time and decides whether $q$ is still a valid clarification
there: grounded in the current text, and asking about something that remains
meaningfully unspecified. Future text is never shown, so $q$ can be a valid
reminder even if the description later supplies the answer. Each question
receives one label,
$
J(q) \in
\{\textsc{Blocker},\textsc{Reminder},\textsc{Irrelevant}\}.
$

PostHoc questions are not evaluated as reminders because they are asked
only after the problem description is complete. At that point, there is
no future authoring context. A question that does not match a registered
blocker is treated as unmatched by the benchmark. The registry and these labels are
used only for evaluation; the interaction and reasoning layers see gold
information only through a resolution that an accepted question has earned.

\subsubsection{Metrics}

Let $Q$ be the questions scored for a task, $B$ its registered blockers, and
$Q_R \subseteq Q$ the questions labeled \textsc{Reminder}. We first report
reminders separately, then exclude them when computing blocker-detection
metrics. Let 
$
Q_{\mathrm{eval}} = Q \setminus Q_R
$
be the remaining questions, and $B_{\mathrm{det}}\subseteq B$ be the
registered blockers covered by at least one question in $Q_{\mathrm{eval}}$.
For PostHoc, $Q_R=\emptyset$, so $Q_{\mathrm{eval}}=Q$.

{\bf \em 1. Reminder Share} measures, among questions that do not cover any
registered blocker, how many are valid given the prefix available
when they are asked. Let $Q_{\mathrm{miss}}\subseteq Q$ denote questions that
match no registered blocker. We define
$
\mathrm{Rem}
=
100\cdot
\frac{|Q_R|}{|Q_{\mathrm{miss}}|}.
$
A blocker-matching question is excluded from both the numerator and
denominator. Thus, Rem quantifies how often an otherwise unmatched interactive
question is still a useful prefix-grounded reminder. It is reported separately
from the blocker-detection metrics below and from the blocker-level
ReminderRate in Section~\ref{sec:rq1}.

{\bf \em 2. Ask Recall} measures the fraction of blockers surfaced
by the non-reminder questions:
$
\mathrm{AskR}
=
100\cdot
\frac{|B_{\mathrm{det}}|}{|B|}.
$
Each blocker is counted at most once regardless of how many questions target it.

{\bf \em 3. Ask Precision} measures the fraction of evaluated questions that
correctly target at least one registered blocker. Let
$
Q_B =
\{q\in Q_{\mathrm{eval}} :
q \text{ covers at least one registered blocker}\}.
$
We define
$
\mathrm{AskP}
=
100\cdot
\frac{|Q_B|}{|Q_{\mathrm{eval}}|}.
$
Because reminders are removed before computing AskP, the metric measures
blocker-targeting precision under the same criterion for interactive and
PostHoc methods.

{\bf \em 4. Ask F1} is the harmonic mean of AskP and AskR:
$
\mathrm{AskF1}
=
2\cdot
\frac{\mathrm{AskP}\cdot\mathrm{AskR}}
{\mathrm{AskP}+\mathrm{AskR}}.
$
It summarizes the trade-off between asking questions targeting to
blockers and covering a large fraction of them.

{\bf \em 5. Tokens at a correct question} measure how early the assistant
surfaces a registered blocker during authoring. We compute the mean prefix
length, in tokens, over questions in $Q_B$. For a question generated at token
$t$, the corresponding prefix contains $t{+}1$ tokens. PostHoc observes the
completed problem description, so its value corresponds to the full prompt
length. 
A smaller value indicates that the model asks a question about a blocker earlier, using less authoring context.


\vspace{-5pt}
\subsubsection{Empirical Results}

Table~\ref{tab:rq2-question-quality} reports the question-quality results.

{\bf \em 1. Interactive clarification improves issue detection over PostHoc.}
Both IPA variants outperform PostHoc in AskR, AskP, and AskF1 across all three
models. IPA (Token+Fallback) improves AskF1 from 44.4 to 59.2 for DeepSeek,
48.1 to 60.1 for Kimi, and 40.2 to 63.2 for Grok, corresponding to gains of
12.0--23.0 percentage points. The same pattern holds for IPA (Chunked), indicating that intermediate
clarification is associated with higher blocker-detection performance than
clarification performed only after the completed description under this
evaluation protocol.

{\bf \em 2. Token+Fallback provides the highest standalone AskF1.}
IPA (Token+Fallback) achieves the highest standalone AskF1 point estimate
for all three models. Compared with IPA (Chunked), it increases recall for
DeepSeek and Kimi.
These results suggest that combining learned
token-level verification with the safety fallback provides a favorable
precision--coverage balance.


{\bf \em 3. IPA surfaces useful clarifications beyond the registered blocker set earlier, while PostHoc improves final coverage.} Across the interactive methods, 36.1--42.9\% of questions that match no
registered blocker are nevertheless judged valid given the prefix available
when they are asked. These prefix-grounded reminders capture a form of
interactive assistance not reflected in blocker recall alone. Adding a final
PostHoc stage further increases blocker coverage, although its additional
questions reduce blocker-targeting precision. Moreover, IPA asks questions much earlier than
PostHoc: at roughly 438--483 tokens on average, compared with 766--817 tokens
for PostHoc.


\vspace{-6pt}

\begin{boxB}{Finding 4}
Interactive clarification improves both the quality and timeliness of issue
detection. IPA (Token+Fallback) achieves the best standalone AskF1 across all
three models, improving over PostHoc by 12.0--23.0 percentage points while
surfacing blocker-related questions substantially earlier. A final PostHoc
stage increases recall further, but at the cost of lower precision.
\end{boxB}

\begin{table*}[t]
\centering
\footnotesize
\caption{Question quality in issue detection. Rem.\ is the fraction of
questions that miss every blocker and are still valid reminders. Tok is the
mean prefix length, in tokens, at a question that covers a blocker.}
\label{tab:rq2-question-quality}
\tabcolsep 2pt
\vspace{-6pt}
\begin{tabular}{l|ccccc|ccccc|ccccc}
\toprule
& \multicolumn{5}{c|}{DeepSeek-V4-Flash}
& \multicolumn{5}{c|}{Kimi-K2.5}
& \multicolumn{5}{c}{Grok-4-1-fast} \\
\cmidrule(lr){2-6}
\cmidrule(lr){7-11}
\cmidrule(lr){12-16}

Method
  & AskR & AskP & AskF1 & Rem. & Tok
  & AskR & AskP & AskF1 & Rem. & Tok
  & AskR & AskP & AskF1 & Rem. & Tok \\
\midrule

PostHoc
  & 54.8 & 37.3 & 44.4 & -- & 765.8
  & 60.6 & 39.9 & 48.1 & -- & 791.6
  & 69.2 & 28.3 & 40.2 & -- & 817.2 \\

IPA (Chunked)
  & 69.2 & \textbf{51.2} & 58.9 & \textbf{42.9} & \textbf{438.5}
  & 71.5 & 48.1 & 57.5 & 40.2 & \textbf{453.3}
  & 75.0 & 51.5 & 61.1 & 36.1 & \textbf{467.9} \\

IPA (Token+Fallback)
  & 72.1 & 50.2 & \textbf{59.2} & 39.8 & 463.4
  & 74.4 & \textbf{50.4} & \textbf{60.1} & \textbf{42.7} & 477.6
  & 73.1 & \textbf{55.6} & \textbf{63.2} & \textbf{40.1} & 482.6 \\

\quad + PostHoc
  & \textbf{76.0} & 40.9 & 53.2 & -- & --
  & \textbf{76.0} & 43.1 & 55.0 & -- & --
  & \textbf{81.7} & 42.0 & 55.5 & -- & -- \\

\bottomrule
\end{tabular}
\end{table*}

\subsubsection{Accuracy of the Trained Interaction Model}
\label{sec:rq2-controller}

The preceding comparison evaluates each interaction schedule end to end,
combining when verification is invoked with downstream question generation.
We next isolate the {\bf \em trained interaction model} and evaluate how accurately it
places \textsc{Verify} actions when the current prefix contains sufficient
evidence for clarification. This separates the learned timing policy
from the reasoning model and the fallback chunking policy in
Table~\ref{tab:rq2-question-quality}.



{\bf \em Label construction: how do we obtain gold token-level actions?} 
HiL-Bench specifies each blocker but not {\em token-level controller actions}.
For each blocker, a LLM annotator marks two positions: the word of the issue onset and the word at which the evidence is sufficient to call reasoning. Tokens before onset are
labeled \textsc{Wait}, tokens from onset to verification are \textsc{Track},
and the verification token is \textsc{Verify}. Annotated positions must be
word tokens.
We annotate a description 3 times independently and use full agreement
to decide the gold positions, with a referee pass for unresolved cases.
Higher agreement on \textsc{Verify} than on issue onset motivates the
tolerance window used~below.



{\bf \em Split.}
We use a fixed $70{:}30$ split by problem description, yielding 70 training
descriptions (249 blockers, 63,183 tokens) and 30 test descriptions (104 blockers). Because
multiple blockers may share a verification token, the test set contains 96
distinct gold \textsc{Verify} events. 

{\bf \em Why \textsc{Verify} is scored with a window.}
Because \textsc{Verify} is a one-token event, exact-token matching is overly
strict. We therefore allow a tolerance window of 4 tokens before and 12 after
the gold event; neighboring events are a median of 53 tokens apart.
We score \textsc{Verify} by one-to-one event matching within this window,
preferring the closest pair. Unmatched pulses are false positives, so repeated
pulsing receives no extra credit.



\begin{wraptable}{r}{3.1in}
    \centering
    \caption{Trained interaction model on the test split.}
    \label{tab:rq2-controller}
    \footnotesize
    \tabcolsep 3.5pt
    \vspace{-9pt}
    \begin{tabular}{lccc}
        \toprule
        \textbf{Method}
        & \textbf{WaitF1}
        & \textbf{TrackF1}
        & \textbf{VerifyF1} \\
        \midrule
        Chunked schedule ($K{=}2$)
        & 0.827 & -- & 0.119 \\
        Trained interaction model
        & 0.797 & 0.537 & \textbf{0.184} \\
        \bottomrule
    \end{tabular}
\end{wraptable}

\textbf{\em Protocol.}
We replay each test description token by token using only the current prefix.
The controller thus conditions on the text seen so far and its previous action, matching the intended inference setting without access to gold labels.
We compare the trained controller with the chunked schedule
(Eq.~\eqref{eq:chunk}, $K{=}2$), which places \textsc{Verify} only at sentence
checkpoints. This protocol allows us to measure whether the learned token-level timing better matches the annotated checking points than the fixed sentence-based checkpoints.






\textbf{\em Results.}
Table~\ref{tab:rq2-controller} shows that the trained interaction model
better identifies when semantic verification should be triggered than the
fixed chunked schedule. Its \textsc{Verify} F1 reaches 0.184, compared with
0.119 for chunking, a 54.6\% relative improvement. The model also achieves a
\textsc{Track} F1 of 0.537, indicating that it can recognize emerging concerns
before verification is warranted.

Its slightly lower \textsc{Wait} F1 is expected because the trained controller
introduces the additional \textsc{Track} state, causing some tokens that would
otherwise be treated as \textsc{Wait} to be classified as \textsc{Track}. The
main objective is the accurate placement of \textsc{Verify} events, where 
the trained controller achieves a higher Verify-F1 than the fixed
chunked schedule.

\subsection{Question Latency Analysis (RQ3)}
\label{sec:rq3}

RQ2 shows that IPA can surface requirement issues from substantially less
context than PostHoc. RQ3 asks what this delay corresponds to in estimated author-facing time:
once an issue becomes sufficiently evident, how long until IPA returns a
clarification question?


Not every concern can be verified when it first appears. In the checkout
example of Section~\ref{sec:motiv}, allowing guest checkout is a reasonable
requirement by itself. The inconsistency becomes evident only later, when the
prompt additionally requires every order to be associated with an account.
IPA should therefore not be penalized for waiting for needed information to establish the issue. We use the annotated \emph{gold verification
point} from Section~\ref{sec:rq2-controller} as the start of the latency
measurement: the point at which the available prefix contains sufficient
evidence to justify verification.

\subsubsection{Latency Measurement}
For each correctly detected blocker, we measure latency from its gold
verification point to the return of the first clarification question that
correctly covers that blocker.
We decompose this latency into two components.

\emph{Layer~1} measures the authoring delay from the gold verification
point to the verification call that first produces a correct question.
We count the intervening tokens and convert this distance to estimated
typing time. We use the free-form composition rate reported
by Vertanen and Kristensson~\cite{10.1145/2555691}, where participants
composed original text at an average of 48 words per minute. On our test
descriptions, this is about $0.81$ seconds per token.

\begin{wraptable}{r}{3.2in}
\centering
\footnotesize
\caption{Latency on successful blocker detection by
IPA (Token+Fallback). Share is the percentage of detected blockers.}
\label{tab:rq3-latency}
\tabcolsep 3pt
\vspace{-6pt}

\begin{tabular}{l l r r r r}
\toprule
Model & Group & Share & Layer 1 (s) & Layer 2 (s) & Total (s) \\
\midrule

\rowcolor{gray!10}
& Easy & 72.0\% & 4.7 & 1.6 & 6.3 \\
\rowcolor{gray!10}
\multirow{-2}{*}{DeepSeek}
& Hard & 28.0\% & 204.9 & 9.6 & 214.5 \\

\midrule

& Easy & 69.4\% & 3.9 & 11.0 & 14.9 \\
\multirow{-2}{*}{Kimi}
& Hard & 30.6\% & 198.0 & 21.3 & 219.3 \\

\midrule

\rowcolor{gray!10}
& Easy & 65.4\% & 2.7 & 25.2 & 27.9 \\
\rowcolor{gray!10}
\multirow{-2}{*}{Grok}
& Hard & 34.6\% & 187.4 & 39.2 & 226.6 \\

\bottomrule
\end{tabular}
\end{wraptable}
\emph{Layer~2} measures the reasoning latency after verification is invoked.
It is the wall-clock time from sending the current prefix to the reasoning
model until the question is returned. \emph{Total} is the sum
of Layer~1 and Layer~2 for each blocker, and we report the mean of these
per-blocker totals.

We further separate correctly detected blockers according to how much additional text IPA requires after the issue becomes verifiable. Context needed before the gold verification point is not counted as latency. We classify a blocker as \emph{Easy} when the first correct question is raised within 50 additional tokens
($t_b^{q}-t_b^{\mathrm{verify}}\leq 50$), and as \emph{Hard} otherwise. Under our typing-rate estimate, 50 tokens correspond to approximately 41 seconds of additional typing.



\vspace{-5pt}
\subsubsection{Empirical Results}

Table~\ref{tab:rq3-latency} reports the latency result.

{\bf \em 1. Most detected blockers are surfaced soon after they become
verifiable.}
Easy blockers account for 65.4--72.0\% of correctly detected blockers.
Their average latency ranges from 6.3 seconds for DeepSeek to
27.9 seconds for Grok. Layer~1 contributes only 2.7--4.7 seconds, showing that for most
detected blockers, the first correct question is initiated shortly after
sufficient evidence.

\textbf{\em 2. Long-delay cases are dominated by additional authoring time.}
The remaining 28.0--34.6\% of correctly detected blockers require
214.5--226.6 seconds on average. Within this group, the estimated
authoring delay to the successful verification call accounts for
187.4--204.9 seconds, substantially more than the reasoning latency.
Thus, long-delay cases arise primarily because the correct question is
triggered substantially after the issue has become verifiable, rather
than because question generation itself is slow. These long-delay cases also help explain the token-position results in
Table~\ref{tab:rq2-question-quality}. Although most correctly detected blockers
are questioned close to their gold check point, a smaller set requires
substantially more authoring before IPA asks the correct question.


\begin{boxB}{Finding 5}
For 65.4--72.0\% of correctly detected blockers, IPA returns a question
within 6.3--27.9 seconds on average after the issue becomes verifiable. The
remaining long-delay cases take 215--227 seconds, with most of the
delay caused by more authoring before verification rather than~reasoning~latency.
\end{boxB}
\vspace{-9pt}
\section{Threats to Validity}
\label{sec:threats}

\noindent\textbf{External validity.}
Although HiL-Bench covers multiple blocker types in SE tasks and model
families, the results may not generalize to other domains, requirement styles,
or humans. The LLM-as-human study provides controlled authoring
trajectories rather than evidence of human~usability.

\noindent\textbf{Internal validity.}
LLM generation and judgment are stochastic and may vary across runs. We
mitigate this via paired comparisons sharing the same initial prompt,
models, decoding configuration, clarification mechanism, and interaction
budget. Rewind-and-resume approximates revise-and-continue authoring but may
not exactly reproduce human behavior. Automated judgments may also introduce
error; our manual audit shows high agreement with the judgment model.
Verification-point annotations may vary. We mitigate via repeated
annotation and a tolerance window.

\noindent\textbf{Construct validity.}
AcceptRate reflects perceived usefulness to a
simulated author rather than objective question quality. 
AskR, AskP, and
AskF1 depend on the HiL-Bench blocker registry and may miss useful
clarifications, partly addressed via prefix-grounded reminders. RQ3
converts token delay from an annotated verification point to estimated typing
time, so authoring latency is estimated rather than directly observed, while
reasoning latency may depend on networks. We do not use revision correctness as an outcome metric because it conflates
IPA's intervention behavior with the downstream capabilities of the reasoning
model (e.g., a weak one) and simulated author.
\vspace{-3pt}
\section{Related Work}
\label{sec:related}

{\bf \em Requirement Clarification for Code Generation}.
CodeClarQA studies code generation through clarification questions
~\cite{li-etal-2023-python}, while TiCoder resolves user intent through
distinguishing test cases~\cite{fakhoury2024ticoder}.
ClarifyGPT detects ambiguous requirements from behavioral inconsistencies and
asks targeted questions~\cite{mu2024clarifygpt}, whereas HumanEvalComm evaluates
whether models recognize incomplete, ambiguous, and inconsistent requirements
and respond appropriately~\cite{wu2025humanevalcomm}.
Other work detects requirement ambiguity or repairs ambiguous specifications
automatically~\cite{bashir2025ambiguity,jia2025repair}.

Recent work increasingly evaluates clarification in interactive coding.
Ambig-SWE studies whether agents can detect underspecification, ask targeted
questions, and use the answers~\cite{vijayvargiya2026ambigswe}, while
Ask or Assume and HiL-Bench focus on selective clarification and help-seeking
~\cite{edwards2026askassumeuncertaintyawareclarificationseeking,
trinh2026hilbenchhumaninloopbenchmarkagents}.
ClarifyCoder and ClarifyCodeBench further study learned clarification behavior
and clarification quality~\cite{wu2025codelanguagemodelslearn,
fang2026clarifycodebench}.
Unlike these approaches, which start from an already presented task, IPA
targets \emph{within-turn clarification} while the requirement is still being
composed.

{\bf \em Proactive Programming Assistance}.
IPA also relates to mixed-initiative systems, where assistance is from users or system~\cite{horvitz1999mixed}.
Chen et al.\ study proactive coding assistance
~\cite{10.1145/3706598.3714002}, while Kuo et al.\ show that developer
receptivity depends strongly on intervention timing and context
~\cite{10.1145/3742413.3789148}.
Related studies further emphasize context awareness and user control
in IDE-based AI assistance~\cite{sergeyuk2026developer,koshchenko2026configurable}.





\vspace{-3pt}
\section{Concluding Remark}
We introduced IPA, a just-in-time assistant that clarifies requirements while
they are being written. By separating interaction control from semantic
verification, IPA complements post-hoc clarification, particularly by helping
surface missing information. More broadly, IPA points toward a new class of prompt assistants. Beyond clarification, such assistants could provide prompt completion, consistency checking, missing-requirement diagnostics, contextual recommendations, and prompt refactoring, opening a broader research agenda on {\bf intelligent prompt assistant} for agentic SE.


\section{Data Availability Statement}

Data/code is available in {\color{black}{project's website~\cite{ipa-website}}}.

\balance

\bibliographystyle{ACM-Reference-Format}

\bibliography{references,references-new}

@article{robillard2010recommendation,
  author  = {Robillard, Martin P. and Walker, Robert J. and Zimmermann, Thomas},
  title   = {Recommendation Systems for Software Engineering},
  journal = {IEEE Software},
  volume  = {27},
  number  = {4},
  pages   = {80--86},
  year    = {2010},
  doi     = {10.1109/MS.2009.161}
}

@inproceedings{svyatkovskiy2020pythia,
  author    = {Svyatkovskiy, Alexey and Zhao, Ying and Fu, Shengyu and
               Sundaresan, Neel},
  title     = {Pythia: {AI}-Assisted Code Completion System},
  booktitle = {Proceedings of the 25th ACM SIGKDD International Conference
               on Knowledge Discovery \& Data Mining},
  pages     = {2727--2735},
  year      = {2019},
  doi       = {10.1145/3292500.3330699}
}

@inproceedings{yang2024sweagent,
  author    = {Yang, John and Jimenez, Carlos E. and Wettig, Alexander and
               Lieret, Kilian and Yao, Shunyu and Narasimhan, Karthik and
               Press, Ofir},
  title     = {{SWE}-agent: Agent-Computer Interfaces Enable Automated
               Software Engineering},
  booktitle = {Advances in Neural Information Processing Systems},
  volume    = {37},
  year      = {2024},
  doi       = {10.52202/079017-1601}
}

@inproceedings{wang2025openhands,
  author    = {Wang, Xingyao and Li, Boxuan and Song, Yufan and Xu, Frank F.
               and Tang, Xiangru and Zhuge, Mingchen and Pan, Jiayi and
               Song, Yueqi and Li, Bowen and Singh, Jaskirat and
               Tran, Hoang H. and Li, Fuqiang and Ma, Ren and
               Zheng, Mingzhang and Qian, Bill and Shao, Yanjun and
               Muennighoff, Niklas and Zhang, Yizhe and Hui, Binyuan and
               Lin, Junyang and Brennan, Robert and Peng, Hao and Ji, Heng
               and Neubig, Graham},
  title     = {{OpenHands}: An Open Platform for {AI} Software Developers
               as Generalist Agents},
  booktitle = {The Thirteenth International Conference on Learning
               Representations},
  year      = {2025}
}

@inproceedings{kamsties2000taming,
  author    = {Kamsties, Erik and Paech, Barbara},
  title     = {Taming Ambiguity in Natural Language Requirements},
  booktitle = {Proceedings of the 13th International Conference on Software
               and Systems Engineering and their Applications},
  volume    = {2},
  year      = {2000},
  publisher = {Centre d'Etude pour la Maitrise des Systemes et du Logiciel
               (CMSL-CNAM)}
}

@incollection{gervasi2019ambiguity,
  author    = {Gervasi, Vincenzo and Ferrari, Alessio and Zowghi, Didar and
               Spoletini, Paola},
  title     = {Ambiguity in Requirements Engineering: Towards a Unifying
               Framework},
  booktitle = {From Software Engineering to Formal Methods and Tools, and Back:
               Essays Dedicated to Stefania Gnesi on the Occasion of Her
               65th Birthday},
  editor    = {ter Beek, Maurice H. and Fantechi, Alessandro and Semini, Laura},
  series    = {Lecture Notes in Computer Science},
  volume    = {11865},
  pages     = {191--210},
  publisher = {Springer},
  year      = {2019},
  doi       = {10.1007/978-3-030-30985-5_12}
}

@article{mu2024clarifygpt,
  author  = {Mu, Fangwen and Shi, Lin and Wang, Song and Yu, Zhuohao and
             Zhang, Binquan and Wang, Chenxue and Liu, Shichao and
             Wang, Qing},
  title   = {{ClarifyGPT}: A Framework for Enhancing {LLM}-Based Code
             Generation via Requirements Clarification},
  journal = {Proceedings of the ACM on Software Engineering},
  volume  = {1},
  number  = {FSE},
  pages   = {2332--2354},
  year    = {2024},
  doi     = {10.1145/3660810}
}

@inproceedings{jia2025repair,
  author    = {Jia, Haoxiang and Morris, Robbie and Ye, He and
               Sarro, Federica and Mechtaev, Sergey},
  title     = {Automated Repair of Ambiguous Problem Descriptions for
               {LLM}-Based Code Generation},
  booktitle = {Proceedings of the 40th IEEE/ACM International Conference
               on Automated Software Engineering},
  pages     = {367--379},
  year      = {2025},
  doi       = {10.1109/ASE63991.2025.00038}
}

@article{fang2026clarifycodebench,
  author        = {Fang, Zheng and Jin, Dongming and Dong, Yihong and
                   Li, Yongmin and Zhang, Kechi and Jin, Zhi and Li, Ge},
  title         = {{ClarifyCodeBench}: Evaluating {LLM}s on Clarifying
                   Ambiguous Requirements for Code Generation},
  journal       = {arXiv preprint arXiv:2607.00711},
  year          = {2026},
  eprint        = {2607.00711},
  archivePrefix = {arXiv},
  primaryClass  = {cs.SE},
  doi           = {10.48550/arXiv.2607.00711}
}

@misc{trinh2026hilbenchhumaninloopbenchmarkagents,
      title={HiL-Bench (Human-in-Loop Benchmark): Do Agents Know When to Ask for Help?}, 
      author={Tu Trinh and Mohamed Elfeki and Guangze Luo and Kelvin Luu and Nathan Hunt and Ernesto Hernandez and Nandan Marwaha and Yannis Yiming He and Charles Wang and Fernando Carabedo and Alessa Castillo and Bing Liu},
      year={2026},
      eprint={2604.09408},
      archivePrefix={arXiv},
      primaryClass={cs.AI},
      url={https://arxiv.org/abs/2604.09408}, 
}

@dataset{ipa-website,
  url          = {https://interactive-pa.site},
}

@article{10.1145/2555691,
author = {Vertanen, Keith and Kristensson, Per Ola},
title = {Complementing text entry evaluations with a composition task},
year = {2014},
issue_date = {February 2014},
publisher = {Association for Computing Machinery},
address = {New York, NY, USA},
volume = {21},
number = {2},
issn = {1073-0516},
url = {https://doi.org/10.1145/2555691},
doi = {10.1145/2555691},
journal = {ACM Trans. Comput.-Hum. Interact.},
month = feb,
articleno = {8},
numpages = {33}
}

@article{fakhoury2024ticoder,
  author  = {Fakhoury, Sarah and
             Naik, Aaditya and
             Sakkas, Georgios and
             Chakraborty, Saikat and
             Lahiri, Shuvendu K.},
  title   = {{LLM}-Based Test-Driven Interactive Code Generation:
             User Study and Empirical Evaluation},
  journal = {IEEE Transactions on Software Engineering},
  volume  = {50},
  number  = {9},
  pages   = {2254--2268},
  year    = {2024},
  doi     = {10.1109/TSE.2024.3428972}
}

@article{wu2025humanevalcomm,
  author    = {Wu, Jie JW and Fard, Fatemeh H.},
  title     = {{HumanEvalComm}: Benchmarking the Communication Competence
               of Code Generation for {LLM}s and {LLM} Agents},
  journal   = {ACM Transactions on Software Engineering and Methodology},
  volume    = {34},
  number    = {7},
  articleno = {189},
  pages     = {1--42},
  year      = {2025},
  doi       = {10.1145/3715109}
}

@inproceedings{bashir2025ambiguity,
  author    = {Bashir, Sarmad and
               Ferrari, Alessio and
               Khan, Abbas and
               Strandberg, Per Erik and
               Haider, Zulqarnain and
               Saadatmand, Mehrdad and
               Bohlin, Markus},
  title     = {Requirements Ambiguity Detection and Explanation with
               {LLM}s: An Industrial Study},
  booktitle = {2025 IEEE International Conference on
               Software Maintenance and Evolution (ICSME)},
  pages     = {620--631},
  year      = {2025},
  publisher = {IEEE},
  doi       = {10.1109/ICSME64153.2025.00063}
}

@inproceedings{sergeyuk2026developer,
  author    = {Sergeyuk, Agnia and
               Koshchenko, Ekaterina and
               Zakharov, Ilya and
               Bryksin, Timofey and
               Izadi, Maliheh},
  title     = {Developer Needs and Feasible Features for
               {AI} Assistants in {IDE}s},
  booktitle = {2026 IEEE/ACM 48th International Conference on
               Software Engineering: Software Engineering in Practice},
  pages     = {406--416},
  year      = {2026},
  publisher = {Association for Computing Machinery},
  doi       = {10.1145/3786583.3786883}
}

@inproceedings{koshchenko2026configurable,
  author    = {Koshchenko, Ekaterina and
               Stankovic, Jovana and
               Zakharov, Ilya and
               Sergeyuk, Agnia},
  title     = {Configurable {AI} Coding Assistants:
               Designing For Developers Who Like to Be in Control},
  booktitle = {Proceedings of the 34th ACM International Conference
               on the Foundations of Software Engineering Companion},
  pages     = {1823--1831},
  year      = {2026},
  publisher = {Association for Computing Machinery},
  doi       = {10.1145/3803437.3806709}
}

@inproceedings{li-etal-2023-python,
    title = "Python Code Generation by Asking Clarification Questions",
    author = "Li, Haau-Sing (Xiaocheng)  and
      Mesgar, Mohsen  and
      Martins, Andr{\'e}  and
      Gurevych, Iryna",
    editor = "Rogers, Anna  and
      Boyd-Graber, Jordan  and
      Okazaki, Naoaki",
    booktitle = "Proceedings of the 61st Annual Meeting of the Association for Computational Linguistics (Volume 1: Long Papers)",
    month = jul,
    year = "2023",
    address = "Toronto, Canada",
    publisher = "Association for Computational Linguistics",
    url = "https://aclanthology.org/2023.acl-long.799/",
    doi = "10.18653/v1/2023.acl-long.799",
    pages = "14287--14306"
}

@inproceedings{
vijayvargiya2026ambigswe,
title={Ambig-{SWE}: Interactive Agents to Overcome Underspecificity in Software Engineering},
author={Sanidhya Vijayvargiya and Xuhui Zhou and Akhila Yerukola and Maarten Sap and Graham Neubig},
booktitle={The Fourteenth International Conference on Learning Representations},
year={2026},
url={https://openreview.net/forum?id=X2yzXtH4wp}
}

@misc{edwards2026askassumeuncertaintyawareclarificationseeking,
      title={Ask or Assume? Uncertainty-Aware Clarification-Seeking in Coding Agents}, 
      author={Nicholas Edwards and Sebastian Schuster},
      year={2026},
      eprint={2603.26233},
      archivePrefix={arXiv},
      primaryClass={cs.CL},
      url={https://arxiv.org/abs/2603.26233}, 
}

@misc{wu2025codelanguagemodelslearn,
      title={Can Code Language Models Learn Clarification-Seeking Behaviors?}, 
      author={Jie JW Wu and Manav Chaudhary and Davit Abrahamyan and Arhaan Khaku and Anjiang Wei and Fatemeh H. Fard},
      year={2025},
      eprint={2504.16331},
      archivePrefix={arXiv},
      primaryClass={cs.SE},
      url={https://arxiv.org/abs/2504.16331}, 
}

@inproceedings{horvitz1999mixed,
author = {Horvitz, Eric},
title = {Principles of mixed-initiative user interfaces},
year = {1999},
isbn = {0201485591},
publisher = {Association for Computing Machinery},
address = {New York, NY, USA},
url = {https://doi.org/10.1145/302979.303030},
doi = {10.1145/302979.303030},
booktitle = {Proceedings of the SIGCHI Conference on Human Factors in Computing Systems},
pages = {159–166},
numpages = {8},
location = {Pittsburgh, Pennsylvania, USA},
series = {CHI '99}
}

@inproceedings{10.1145/3706598.3714002,
author = {Chen, Valerie and Zhu, Alan and Zhao, Sebastian and Mozannar, Hussein and Sontag, David and Talwalkar, Ameet},
title = {Need Help? Designing Proactive AI Assistants for Programming},
year = {2025},
isbn = {9798400713941},
publisher = {Association for Computing Machinery},
address = {New York, NY, USA},
url = {https://doi.org/10.1145/3706598.3714002},
doi = {10.1145/3706598.3714002},
booktitle = {Proceedings of the 2025 CHI Conference on Human Factors in Computing Systems},
articleno = {881},
numpages = {18},
location = {
},
series = {CHI '25}
}

@inproceedings{10.1145/3742413.3789148,
author = {Kuo, Nadine and Sergeyuk, Agnia and Chen, Valerie and Izadi, Maliheh},
title = {Developer Interaction Patterns with Proactive AI: A Five-Day Field Study},
year = {2026},
isbn = {9798400719844},
publisher = {Association for Computing Machinery},
address = {New York, NY, USA},
url = {https://doi.org/10.1145/3742413.3789148},
doi = {10.1145/3742413.3789148},
booktitle = {Proceedings of the 31st International Conference on Intelligent User Interfaces},
pages = {349–362},
numpages = {14},
location = {
},
series = {IUI '26}
}

\end{document}